\documentclass[aps,prd,twocolumn,superscriptaddress,showpacs,nofootinbib]{revtex4-1}

\usepackage[utf8]{inputenc}

\usepackage{diagbox}

\usepackage{mathtools}
\usepackage{amsfonts}
\usepackage{mathrsfs}
\usepackage{bbm}
\usepackage{slashed}

\usepackage{graphicx}
\usepackage{color}
\usepackage{array}
\usepackage{esint}
\usepackage{placeins}
\usepackage{booktabs}
\usepackage{makecell}
\usepackage{epstopdf}
\usepackage[caption=false]{subfig}

\usepackage{xspace}
\usepackage{siunitx}
\usepackage{hyperref}
\usepackage[nameinlink]{cleveref}
\usepackage{appendix}

\usepackage{xifthen}
\usepackage{xcolor}
\hypersetup{
	colorlinks,
	linkcolor={red!75!black},
	citecolor={blue!75!black},
	urlcolor={blue!75!black}
}

\usepackage{comment}
\usepackage{dsfont}
\usepackage{tensor}
\usepackage{tabularx}
\usepackage{orcidlink}

\begin{document}


\title{From Magnetic to Inverse Magnetic Catalysis: The Interplay of Quark and Gluon Mass Generation in Magnetic Fields}


\author{Fei Gao\,\orcidlink{0000-0001-5925-5110}}
\email{fei.gao@bit.edu.cn}
\affiliation{School of Physics, Beijing Institute of Technology, 100081 Beijing, China}

\author{Kairen Huang}
\affiliation{School of Physics, Beijing Institute of Technology, 100081 Beijing, China}

\author{Yi Lu}
\email{qwertylou@pku.edu.cn}
\affiliation{School of Physics, Peking University, 100871 Beijing, China}

\author{Yuxin Liu}
\email{yxliu@pku.edu.cn}
\affiliation{School of Physics, Peking University, 100871 Beijing, China}
 \affiliation{Center for High Energy Physics, Peking University, 100871 Beijing, China}
\affiliation{Collaborative Innovation Center of Quantum Matter, Beijing 100871, China}



\date{\today}

\begin{abstract}

We analyze the effects of the magnetic field on the quark and gluon propagators within the functional QCD framework. By solving the coupled Dyson–Schwinger equations for the quark and gluon propagators, we find that the quark mass is generally enhanced in the presence of a magnetic field, leading to magnetic catalysis of the chiral condensate. Meanwhile, the magnetic field also induces an increase in the gluon screening mass. The enhancement of the gluon screening mass suppresses the quark–gluon interaction and thereby weakens the strength of dynamical chiral symmetry breaking, establishing a competing mechanism against magnetic catalysis. In particular, this enhancement of the gluon screening mass becomes dominant near the chiral phase transition, which in turn gives rise to inverse magnetic catalysis.
\end{abstract}


\maketitle



%

\section{Introduction}
The heavy ion collisions create extremely strong magnetic field background  by  the   non central collision between the two charged nuclei~\cite{Rafelski:1975rf,Skokov:2009qp,Voronyuk:2011jd}. The presence of  the magnetic field  brings in rich  phenomena, making it necessary to incorporate magnetic field effects into QCD theory in order to fully understand the hot QCD matter created in heavy-ion collision experiments.  This has attracted significant theoretical interest,  however,   there is still lack of comprehensive understanding for the impact of the finite magnetic field on the  basic feature of QCD, i.e.,  the chiral symmetry breaking.  
In vacuum, it is widely accepted that the magnetic field enhances chiral symmetry breaking. Theoretical calculations, including lattice QCD simulations and effective models, confirm that the chiral condensate increases with the magnetic field strength—a phenomenon known as magnetic catalysis (MC)~\cite{Gusynin:1999pq,Leung:2005xz,Cohen:2007bt,Boomsma:2009yk,DElia:2011koc,Fukushima:2012xw,Gatto:2012sp,Gorbar:2013uga,Miransky:2015ava,Ding:2026qzu}.  However, near the chiral phase transition temperature, lattice QCD simulations reveal that the chiral condensate decreases with increasing magnetic field strength, and the chiral phase transition temperature is likewise reduced—a phenomenon known as inverse magnetic catalysis (IMC)~\cite{Endrodi:2024cqn,Bali:2011qj,Bali:2012jv,Bali:2014kia,Bruckmann:2013oba,Ding:2020inp,DElia:2021yvk,Ding:2022tqn,Ding:2025pbu}.  

Several attempts have been made to understand the IMC within effective models~\cite{Andersen:2014xxa,Hattori:2023egw,Fukushima:2012kc,Ferreira:2013tba,Ferreira:2014kpa,Andersen:2014oaa,Ayala:2015bgv, Yu:2014xoa,Yu:2014sla,Li:2016gfn,Zhu:2023aaq,ZHU:2024gxy,Mao:2016fha,Mao:2025toi}.  However, it is generally difficult to reproduce the lattice results: at low temperatures, QCD matter exhibits MC, whereas near the phase transition temperature at high temperatures, IMC emerges. In particular, it has been found that incorporating a suppression of the effective coupling induced by the finite magnetic field into the calculation can reproduce the IMC phenomenon. Therefore, a self-consistent determination of the QCD running behavior may provide a satisfactory explanation for the IMC effect.

 Functional QCD approaches, and here specifically, the Dyson-Schwinger equations (DSEs) approach, provide a platform for the quantitative description of the QCD running behavior from the ultraviolet to the infrared regime, based on the fundamental degrees of freedom of the theory—quarks and gluons. Numerous efforts have been made within the DSE framework in a magnetic field to study the MC and IMC effects~\cite{Watson:2013ghq,Mueller:2014tea,Miransky:2015ava,Mueller:2015fka,Ding:2025zqu}.  In particular,  in Ref.~\cite{Mueller:2015fka}, it has been  argued  that the gluon screening mass together with the suppression of the QCD coupling lead to the IMC phenomenon. 
 In the present work, we then present a self-consistent calculation of the magnetic field dependence of the quark and gluon propagators within the minimal QCD scheme of the DSEs. We find that the results are consistent with lattice QCD observations, namely, MC at low temperatures and IMC at high temperatures.

 The manuscript is organised as follows: we firstly generalize the minimal QCD scheme  within the magnetic field in \Cref{sec:mDSEs}. In specific, the magnetic field dependence of the quark propagator and its impact on gluon propagator via the quark loop are considered.   In \Cref{sec:numerics}, we show results on the quark and gluon   mass  scale  under the magnetic field,  furthermore,   discussions  on the  magnetic catalysis at low temperature and  the Inverse Magnetic catalysis at high temperature.   
 Finally, in \Cref{sec:sum} we summarise the results  and make some conclusions and outlooks.
 
\section{The framework of DSEs with the presence of  the magnetic field}\label{sec:mDSEs}
In this section, we present the generalization of the minimal QCD truncation scheme to include the presence of a magnetic field. The central ingredient is the non-perturbative generalization of the Schwinger propagator, which is computationally simpler than the Ritus formalism for numerical calculations. We demonstrate how to incorporate the non-perturbative Schwinger propagator into the gap equation within a relatively general setting for the interaction kernel, and we further discuss its impact on the gluon propagator through the quark loop diagram.
\subsection{Quark propagator in the magnetic field}
With the presence of  a constant magnetic field,  the quark propagator  is no longer the eigenfunction of the covariant derivative operator in the momentum representation. Instead, the Ritus basis based on the Hermite funciton is applied since one can decompose the  kinetic term as:
\begin{eqnarray}
&&D\!\!\!/+m=(\partial_x \gamma_1 +iq_f B x \gamma_2+i\bar{p}\!\!\!/+m_0)\\
=&&\Sigma^+(ip\!\!\!/_\parallel+m_0)+\Sigma^-(ip\!\!\!/_\parallel+m_0)\notag\\
&&+I_+(\partial_{{x}_{p_y}}+q_f B{x}_{p_y})+I_-(\partial_{{x}_{p_y}}-q_f B{x}_{p_y}),\notag
\end{eqnarray}
with the electromagnetic gauge field as $A^{\rm EM}_\mu=(0, 0, q_f Bx,0)$ and $I^{\pm}=\frac{\gamma_1\pm\gamma_2}{2} $, $\Sigma^\pm=\frac{1\pm\Sigma_3}{2}$ and $\Sigma_3=i\gamma_1\gamma_2$. The properties of the Hermite function together with the Dirac structure allows one to find a finite dimensional  complete basis, which is the Ritus basis~\cite{Ritus:1972ky,Ritus:1978cj}. 
The Ritus formalism is closely related to the Landau level  through the order of the Hermite function.  

For a  tree level propagator under the constant magnetic field, one can further obtain the Schwinger propagator by  applying a proper time formalism to the propagator in Ritus formalism, making it possible to sum up  the Hermite function~\cite{Schwinger:1951nm, Chodos:1990vv}.  The Schwinger propagator has great advantages in numerical calculations as it can be expressed in momentum representation besides of a factor which is called the Schwinger phase~\cite{Gusynin:1999pq,Watson:2013ghq,Ding:2025zqu}.
Moreover, 
as has been illustrated in Refs.~\cite{Gusynin:1999pq,Miransky:2015ava},  the Schwinger phase  can be completely factored  out  from the gap equation if the interaction kernel contains only the transferred momentum. The price is that the relation between the inverse quark propagator $S^{-1}(x,y)$  and the propagator $S^{}(x,y)$ becomes more complicated as they have the same dependence on the Schwinger phase which reads as:
\begin{eqnarray}\label{eq:SchPropa}
&&S^{-1}(x,y)=e^{i\Phi(x,y)}{\bar S}(x-y),\notag\\
&&S^{}(x,y)=e^{i\Phi(x,y)}S^{}(x-y), 
\end{eqnarray}
with $\Phi(x,y)$ the Schwinger phase  as $$\Phi(x,y)=- q_f B (x-y)_\mu A^{\rm EM}_\mu(x+y).$$  

The functions ${\bar S}(x-y)$ and $S^{}(x-y)$ in Eq.~\ref{eq:SchPropa} can be further expressed in momentum representation as:
\begin{eqnarray}\label{eq:SchPropa1}
&&{\bar S}(x-y)=\int \frac{d^4p}{(2\pi)^4}{\bar S}(p)e^{ip(x-y)},\, \notag\\
&&{ S}(x-y)=\int \frac{d^4p}{(2\pi)^4}{  S}(p)e^{-ip(x-y)}, 
\end{eqnarray}
The factoring out of the Schwinger phase leads to a sophisticated result that  the gap equation  in constant magnetic field  in terms of ${\bar S}(p)$  and ${S}(p) $  is precisely the same as the conventional gap equation without magnetic field, which reads as:

\begin{eqnarray}
&&{\bar S} (p) =S_0(p)+\Sigma(p),\notag\\
&&\Sigma(p)=T\sum_{n\in \mathbbm{Z}} \int \frac{d^3 q}{(2 \pi)^3}\frac{4}{3} g_s \gamma_\mu S(q)\Gamma_{\nu}(k) D_{\mu\nu}(k)\,,
		\label{eq:QuarkDSE2}
	\end{eqnarray}
where the momentum is parametrized as  $p=(p_\parallel, p_\perp)$ with $p_\parallel=(  (2 m+1)\pi T+i \mu,p_z)$ and $p_\perp=(p_x,p_y)$. 
Note that  ${\bar S} (p) $ is no longer the inverse of $ S(p)$ as discussed above. Nevertheless, one can still parameterize ${\bar S} (p) $ and $S(p)$ in the same Dirac structure basis:
\begin{eqnarray}
{\bar S} (p)&&=\Sigma^+(M^+(p)  + Z^+_\parallel(p)  \slashed{p}_\parallel) + Z_\perp(p)  \slashed{p}_\perp \notag\\
&&+\Sigma^-(M^-(p)  + Z^-_\parallel(p)  \slashed{p}_\parallel) ,\\
 S (p)&&=\Sigma^+(W_{M^+}(p)  +W_{ Z^+_\parallel}(p)  \slashed{p}_\parallel) + W_{Z_\perp}(p)  \slashed{p}_\perp \notag\\
&&+\Sigma^-(W_{M^-}(p)  + W_{Z^-_\parallel}(p)  \slashed{p}_\parallel).
		\label{eq:propa}
	\end{eqnarray}
There contains more Dirac structures that represent the splitting of spin up and down fermions due to the presence of the magnetic field.  The additional structures in terms of $\Sigma_3$ are related to various anomalous phenomena, for example, $M^+-M^-$  and $ Z^+_\parallel-Z^-_\parallel$ are related to the Pauli term and thus the anomalous magnetic moment~\cite{Ferrer:2015wca}, and    contribute to the anomalous Zeeman effect~\cite{Ferrer:2008dy,Ding:2025zqu}.  Moreover, both terms are also related to the chiral magnetic/separation effect~\cite{Gao:2026bju}.

As mentioned above, ${\bar S} (p) $  defined in Eq.~\ref{eq:SchPropa1} is no longer the inverse quark propagator, which means that  the coefficients of $S(p)$ and  ${\bar S} (p) $ do not follow the conventional inverse relation.  Instead,  their relation can be expressed through the Schwinger propagator formula. One can still parametrize ${\bar S} (p) $  as in Eq.~\ref{eq:propa} and then the coefficients in the quark propagator $S (p)$ can be expressed as:

\begin{eqnarray}\label{eq:propa}
&&W_{ Z^+_\parallel}(p) =\int ds e^{-s\frac{ \Delta^+(p)}{ Z^2_\perp(p)  }-\frac{p_\perp^2}{q_f B}{\rm tanh}[q_f B s ]}\notag\\
&&\times \left\{(1-{\rm tanh}[q_f B s])\frac{Z^-_\parallel(p) }{Z^2_\perp(p)}+s\frac{\Delta^-}{Z^4_\perp(p)} M^-(p)\right\}\,,\notag\\
&&W_{ Z^-_\parallel}(p) =\int ds e^{-s\frac{ \Delta^+(p)}{ Z^2_\perp(p)  }-\frac{p_\perp^2}{q_f B}{\rm tanh}[q_f B s ]}\notag\\
&&\times \left\{(1+{\rm tanh}[q_f B s])\frac{Z^+_\parallel(p) }{Z^2_\perp(p)}-s\frac{\Delta^-}{Z^4_\perp(p)} M^+(p)\right\}\,,\notag\\
&&W_{ Z_\perp}(p) =\int ds e^{-s\frac{ \Delta^+(p)}{ Z^2_\perp(p)  }-\frac{p_\perp^2}{q_f B}{\rm tanh}[q_f B s ]}\notag\\
&&\times \left\{(1-{\rm tanh}^2[q_f B s])\frac{1 }{Z_\perp(p)}\right\}\,,\notag
\end{eqnarray}
\begin{eqnarray}
&&W_{ M^+}(p) =\int ds e^{-s\frac{ \Delta^+(p)}{ Z^2_\perp(p)  }-\frac{p_\perp^2}{q_f B}{\rm tanh}[q_f B s ]}\notag\\
&&\times \left\{(1-{\rm tanh}[q_f B s])\frac{M^+(p) }{Z^2_\perp(p)}-s\frac{\Delta^-}{Z^4_\perp(p)} Z^-_\parallel(p)p^2_\parallel \right\}\,,\notag\\
&&W_{ M^-}(p) =\int ds e^{-s\frac{ \Delta^+(p)}{ Z^2_\perp(p)  }-\frac{p_\perp^2}{q_f B}{\rm tanh}[q_f B s ]}\notag\\
&&\times \left\{(1+{\rm tanh}[q_f B s])\frac{M^+(p) }{Z^2_\perp(p)}+s\frac{\Delta^-}{Z^4_\perp(p)} Z^+_\parallel(p)p^2_\parallel \right\}\,,\notag
	\end{eqnarray}
	with $$\Delta^+(p)=p_\parallel^2 Z^+_\parallel(p)Z^-_\parallel(p)  +M^+(p)M^-(p),$$ and $$\Delta^-(p)= Z^+_\parallel(p)  M^-(p) - M^+(p)  Z^-_\parallel(p). $$
Putting these expressions into the gap equation  together with the information of gluon propagator and quark gluon vertex, one can solve the gap equation and obtain the quark propagator in the magnetic field.

   \begin{figure*}[t]
  \includegraphics[width=0.92\columnwidth]{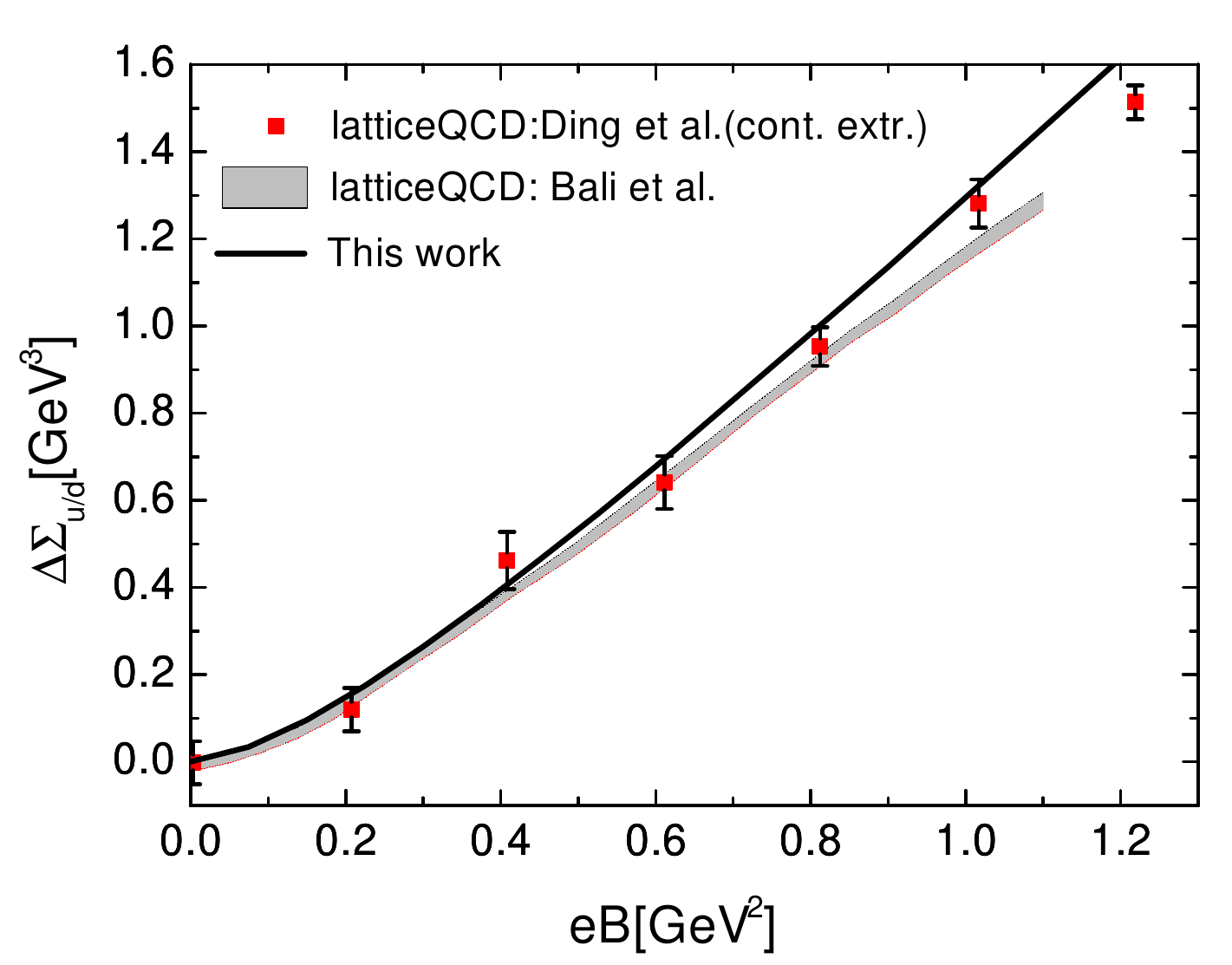}
    \includegraphics[width=0.92\columnwidth]{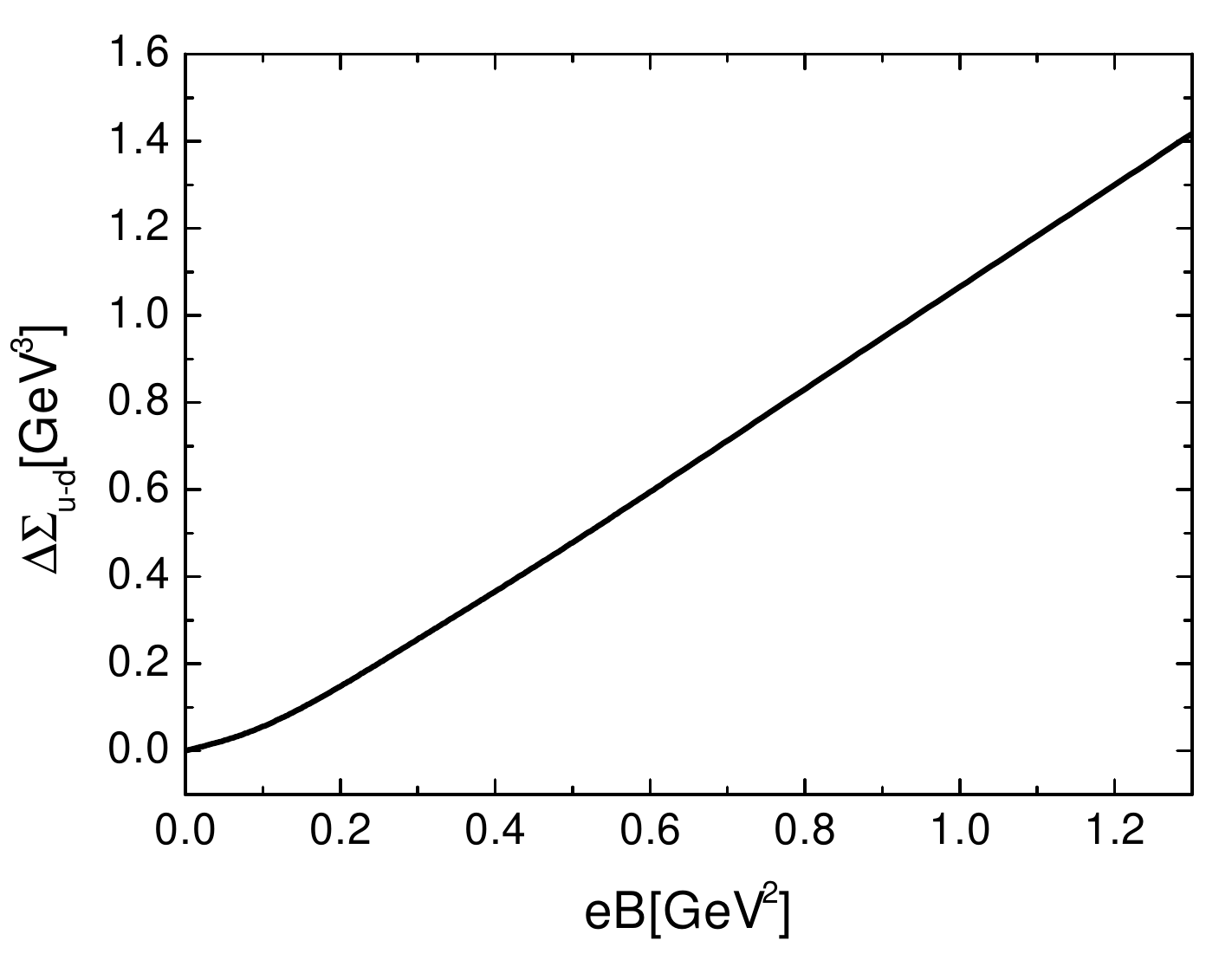}
  \caption{ The magnetic field dependence of the subtracted chiral condensate in vacuum in comparison to the lattice results~\cite{Bali:2012zg,Ding:2026qzu}. The left panel is the average of the up and down quark, and the right panel is the difference between up and down quark. }\label{fig:MC}
\end{figure*}

\subsection{The correction of gluon propagator in the magnetic field}

The gluon propagator can be obtained via the gluon gap equation, which  we take the following difference form of DSE~\cite{Gao:2020qsj, Lu:2023mkn,Gao:2026pfr}:
\begin{align}
    G_{\mu\nu,\vec{v}}^{-1}(k) = G_{\mu\nu,\vec{v}_0}^{-1}(k) + \Delta \Pi^{\vec{v},\vec{v}_0}_{A,\mu\nu}(k),
    \label{eq:diffDSEforGluonProp}
\end{align}
and here the difference is for the vanishing and the  finite magnetic field:
\begin{gather}
    \vec{v}_0 = \Big( N_f , m_f, T, \mu_q, 0 \Big), \nonumber\\
    \vec{v} = \Big( N_f, m_f, T, \mu_q, q_fB \Big).
\end{gather}

The self-energy in the difference DSE in Eq.~\eqref{eq:diffDSEforGluonProp} for the gluon propagator is then given by the  quark vacuum polarization 
\begin{align}
    &\Delta \Pi^{\vec{v},\vec{v}_0}_{A,\mu\nu}(k) \nonumber\\
    &= - \int_q \, \operatorname{tr} \Big[ g_s\gamma_\mu S_f^m(\tilde q) \Gamma^m_{\nu}(\tilde q, \tilde q - k) S_f^m(\tilde q - k) \Big]\bigg|^{m= |q_fB|}_{ m=0},
    \label{eq:QuarkLoop}
\end{align}
with $S_f^m(\tilde q)$ the quark propagator calculated from Eq.~\ref{eq:QuarkDSE2} with the magnetic field strength as  $m=|q_f B|$ and for up and down quark, $q_f=2/3 e$ and $q_f=-1/3 e$ respectively. The self-energy is calculated in Landau gauge and hence the projected self energy is solved as:
\begin{align}
 P^{}_{\mu\nu} \Delta \Pi^{\vec{v},\vec{v}_0}= \Delta \Pi^{\vec{v},\vec{v}_0}_{A,\mu\nu}(k),\quad   P^{}_{\mu\nu}   = \delta_{\mu\nu} -  \frac{k_\mu k_\nu}{k^2}.
\end{align}
The respective gluon propagator is thus written as:
\begin{align}
    G_{\mu\nu,\vec{v}}^{}(k) =  P^{}_{\mu\nu}    G_{\vec{v}}^{}(k)=  \frac{P_{\mu\nu} }{G_{\vec{v}_0}^{-1}(k)+ \Delta \Pi^{\vec{v},\vec{v}_0}},
    \label{eq:GluonProp}
\end{align}
with $G_{\mu\nu,\vec{v}_0}^{-1}(k)$ the gluon propagator at vanishing magnetic field  with its temperature and chemical potential dependence  being  calculated in the similar difference scheme  that has been introduced in the previous studies~\cite{Chen:2026pqz,Gao:2026pfr}. 
\subsection{The quark gluon vertex in the magnetic field}

At last,  to solve the DSEs of quark and gluon propagator, one needs to apply an Ansatz for the quark gluon vertex, and here we apply the minimal QCD scheme ~\cite{Lu:2023mkn,Zheng:2023tbv}  but with some further corrections to fit the requirements of the gap equation in magnetic field. The two dominant Lorentz structures are chosen as in minimal QCD scheme, i.e. the Dirac and Pauli term as:
 \begin{equation}
	\Gamma_{\mu}(k)=\gamma_{\mu}\lambda_1(k)+\sigma_{\mu\nu}k_\nu\lambda_4(k)
 \end{equation}
with
 \begin{equation}
	\lambda_1(k)=F(k^2)Z_\perp(k) \, ,\quad \lambda_4(k)=\eta F(k^2)\frac{\partial M(k^2)}{\vec{k}^2}.
 \end{equation}
 $M(k^2)$ the average quark mass function for $+$ and $-$ components at zeroth Matsubara frequency and $F(k^{2})$ is the ghost dressing function. The derivative of $M(k^2)$ is originally the difference form with momentum $p$ and $q$ separately as in  Ref.~\cite{Lu:2023mkn}. Here we use the derivative because as described before, it is only possible to apply the gap equation at finite magnetic field in Eq.~\ref{eq:QuarkDSE2}  when the vertex and the gluon propagator contain solely the transferred momentum $k=p-q$.  The parameter $\eta=2.68$ is to compensate this modification.  
This then completes the framework of Dyson-Schwinger equations in the constant magnetic field. In the following section, we will represent the results of the quark and gluon propagator and elaborate their relation to the magnetic catalysis and Inverse magnetic catalysis.

\section{Numerical results}\label{sec:numerics}
Effective model studies have confirmed that the quark mass is always enhanced by the magnetic field, leading to magnetic catalysis of the chiral condensate. However, as has been shown by perturbative studies, the gluon screening mass is also enhanced by the magnetic field due to the quark loop contribution to the gluon propagator's self-energy~\cite{Huang:2022fgq,Mueller:2015fka}.  When one considers the effective running coupling of the theory—for instance, the Taylor coupling~\cite{Blossier:2011tf,Blossier:2012ef,Binosi:2016nme} which can be expressed as $\alpha(\mu) F^2(k)Z_g(k)$  with $\alpha(\mu)$ the coupling constant at renormalization scale $\mu$, and $F,Z_g$ the ghost, gluon propagators' dressing functions.  It becomes clear that the enhancement of the gluon mass suppresses the gluon dressing function, thereby leading to a suppression of the effective running coupling. We first illustrate this competing mechanism  and then verify the emergence of inverse magnetic catalysis near the chiral phase transition.
\subsection{Magnetic catalysis in vacuum}

For comparison, we define the subtracted and normalized condensate as introduced in Ref.~\cite{Bali:2011qj} as:
\begin{eqnarray}\label{eq:conden}
\Delta\Sigma_{u/d}&&=\frac{\Delta\Sigma_u(eB)+\Delta\Sigma_d(eB)}{2}\notag\\
&&=\frac{m_f}{m_\pi^2 f_{\pi}^2}[\langle\bar{\psi}\psi \rangle_u(eB)-\langle\bar{\psi}\psi \rangle_u(eB=0)\notag\\
&&+\langle\bar{\psi}\psi \rangle_d(eB)-\langle\bar{\psi}\psi \rangle_d(eB=0)],
\end{eqnarray}
with $m_{f=u/d}=2$ MeV renormalized at $\mu=40$ GeV, $m_\pi=135$ MeV and $f_\pi=86$ MeV as  the pion decay constant in chiral limit.

Besides, the magnetic field response for up and down quark is different. Since the electric charge of up quark is twice of the charge of down quark, and hence, the magnetic effect on up quark is  generally larger than that on down quark. Such a difference can be quantified by the following difference of the $u/d$ averaged chiral condensate as:
\begin{eqnarray}\label{eq:conden1}
\Delta\Sigma_{u-d}&&=\Delta\Sigma_{u}(eB)-\Delta\Sigma_{d}(eB)\notag\\
&&=\frac{2m_f}{m_\pi^2 f_{\pi}^2}[\langle\bar{\psi}\psi \rangle_u(eB)-\langle\bar{\psi}\psi \rangle_d(eB)].
\end{eqnarray}

We first examine the magnetic field dependence of the $u/d$ averaged  chiral condensate defined in Eq.~\ref{eq:conden}. As depicted in Fig.~\ref{fig:MC}, the averaged chiral condensate increases with the magnetic field strength, exhibiting magnetic catalysis. Our results are in good agreement with previous lattice QCD calculations, demonstrating the validity of our framework. This magnetic catalysis behavior arises from the direct enhancement of the quark mass in the presence of the magnetic field. In particular, the enhancement is proportional to the quark mass function: for a larger quark mass, the enhancement is more pronounced. This direct enhancement persists at finite temperature if one neglects the magnetic field dependence in the quark loop feedback to the gluon propagator's self-energy, as in Eq.~\ref{eq:QuarkLoop}.

 In the following, we show that once the quark loop feedback effect is included, the enhancement of the gluon mass scale gives rise to inverse magnetic catalysis. Note that the above vacuum calculation already includes the quark loop correction in Eq.~\ref{eq:QuarkLoop}. However, this correction is weak compared to the direct enhancement of the quark mass function, and hence, the correction in the gluon propagator can only compete with the direct enhancement near the chiral phase transition, where the quark mass function is relatively small.

   \begin{figure*}[t]
  \includegraphics[width=0.89\columnwidth]{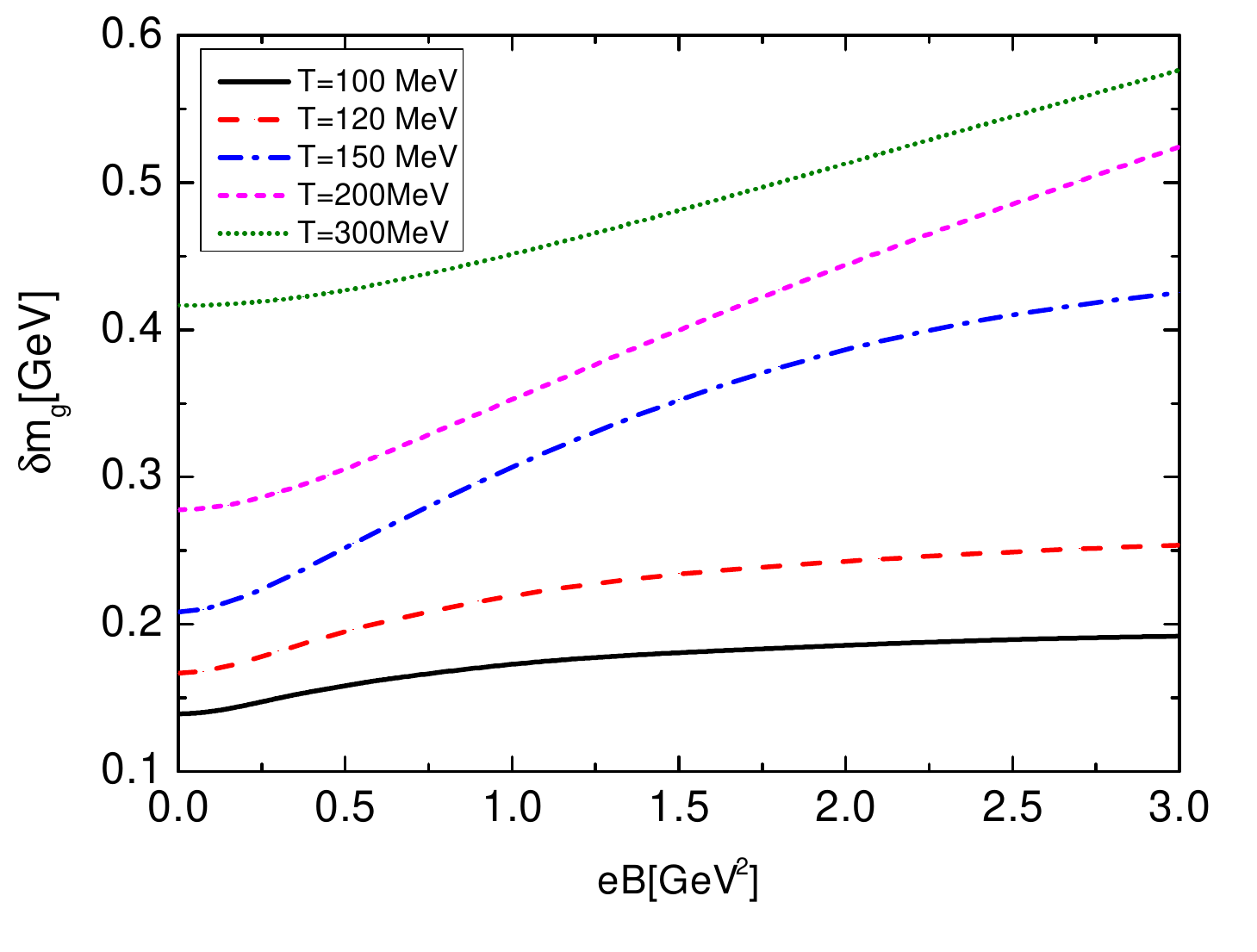}
    \includegraphics[width=0.89\columnwidth]{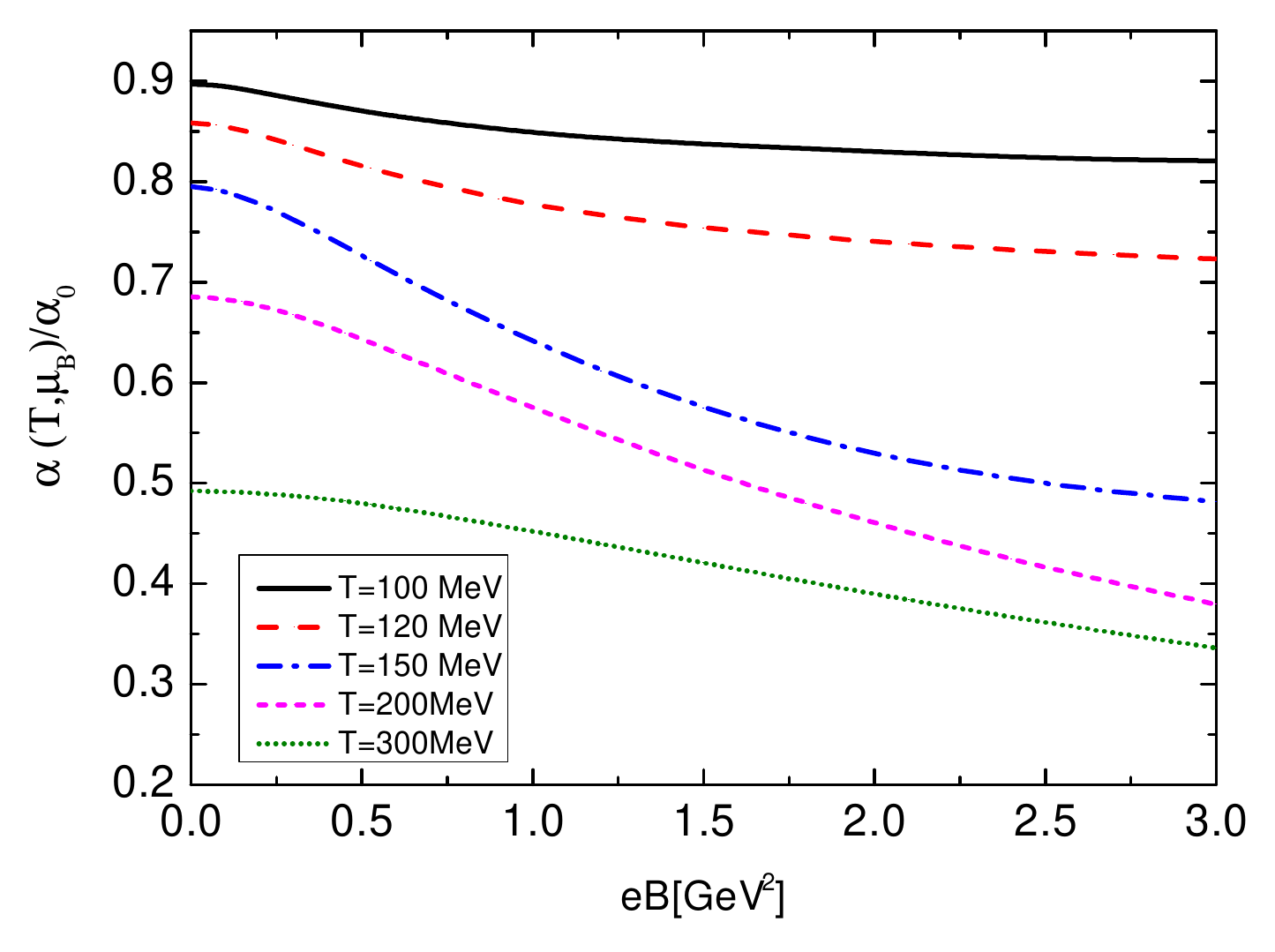}
  \caption{The mass scale of gluon from the quark loop correction in the self energy of gluon propagator and the corresponding QCD coupling strength  estimated by the Taylor coupling in infrared limit, $\alpha(T,\mu_B,eB)=\alpha(\mu) F^2(k)Z_g(k;T,\mu_B,eB)|_{k=0}$. Note that here   the $T,\mu_B, eB$ dependence of the ghost dressing is neglected, which is a valid approximation in the $T,\mu_B$ regime we considered here~\cite{Gao:2020qsj,Lu:2023mkn}, and moreover, is valid for the  $eB$ dependence as the ghost and quark has no direct coupling and  interacts only through the gluon.   }\label{fig:quarkloop}
\end{figure*}

 \begin{figure}[t]
  \includegraphics[width=0.95\columnwidth]{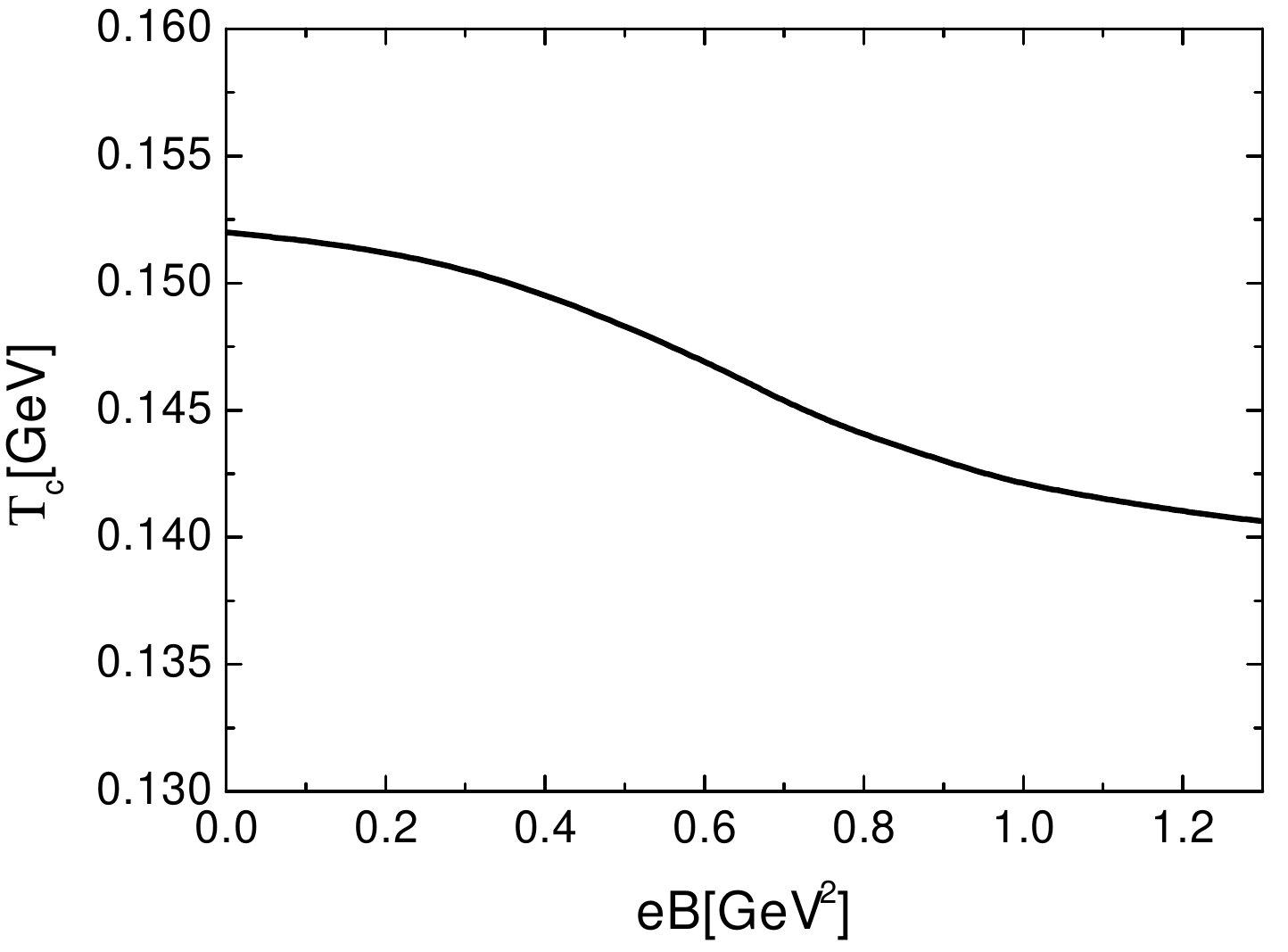}
  \caption{The magnetic field dependence of the pseudo chiral phase transition temperature $T_c$ for vanishing chemical potential obtained from the averaged chiral condensate  for up and down quark.  }\label{fig:Tc}
\end{figure}
\subsection{The inverse magnetic catalysis near chiral phase transition}
 As pointed out in Ref.~\cite{Mueller:2015fka,Huang:2022fgq}, not only the quark mass function, but also the gluon mass function is enhanced by the magnetic field. Such an enhancement comes from the quark loop in the gluon propagator's self energy as in Eq.~\ref{eq:QuarkLoop}.  To quantify the quark loop correction on the gluon mass scale together with the $T,\,\mu_B$ correction from the gluon self interaction, we define the difference of  the  effective gluon mass scale   between the vacuum and finite $T,\,\mu_B,\,eB$ as:
 \begin{align}\label{eq:mg}
 \delta m_g^2
 ={P_{\mu\nu} [D_{\mu\nu,\vec{v}=(T,\mu_B,eB)}^{-1}(k)-D_{\mu\nu,(0,0,0)}^{-1}(k)}]  \bigg |_{k=0}.
 \end{align}
 
 Such an additional mass scale in gluon propagator suppresses the effective quark gluon coupling. This can be directly measured by the Taylor coupling as
  $$\alpha(T,\mu_B,eB)=\alpha(\mu) \left [F^2(k)Z_g(k;T,\mu_B,eB)\right ]\bigg |_{k=0}.$$
 Therefore, the enhancement of the gluon mass scale leads to the suppression of the effective coupling and may offer an explanation of the inverse magnetic catalysis which stands for a suppression of chiral condensate with the presence of the magnetic field.  
 
    \begin{figure*}[t]
  \includegraphics[width=0.9\columnwidth]{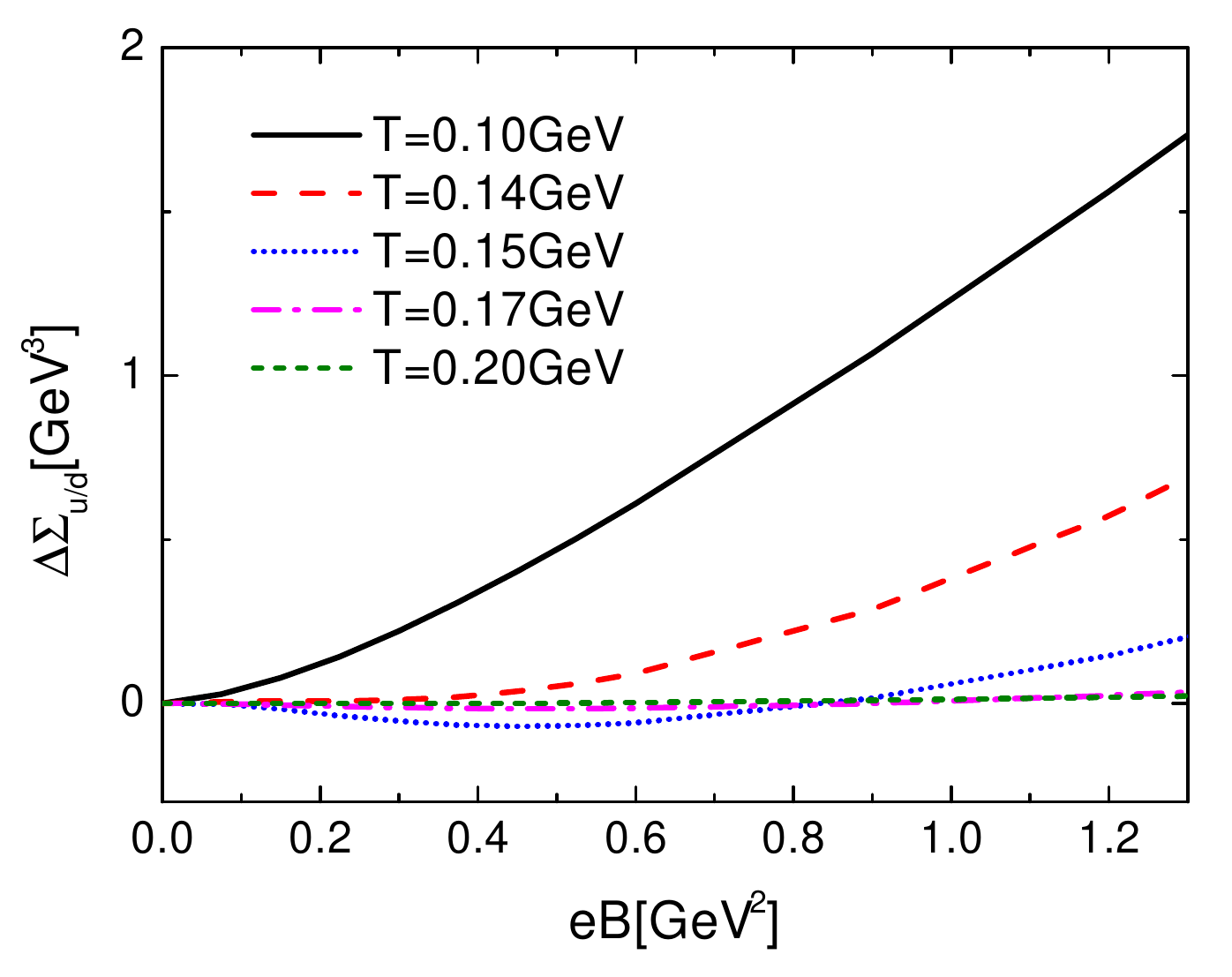}
    \includegraphics[width=0.9\columnwidth]{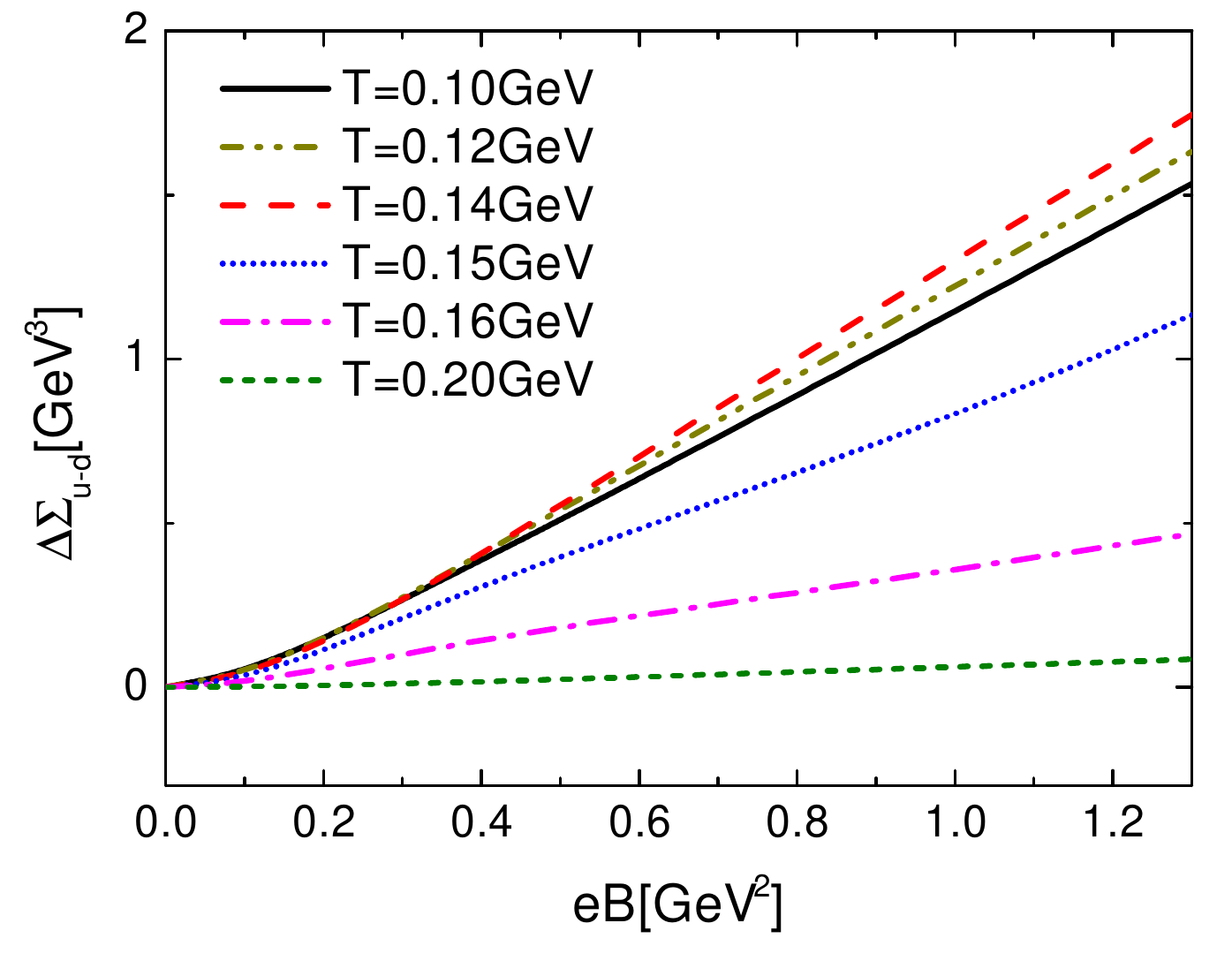}
      \includegraphics[width=0.9\columnwidth]{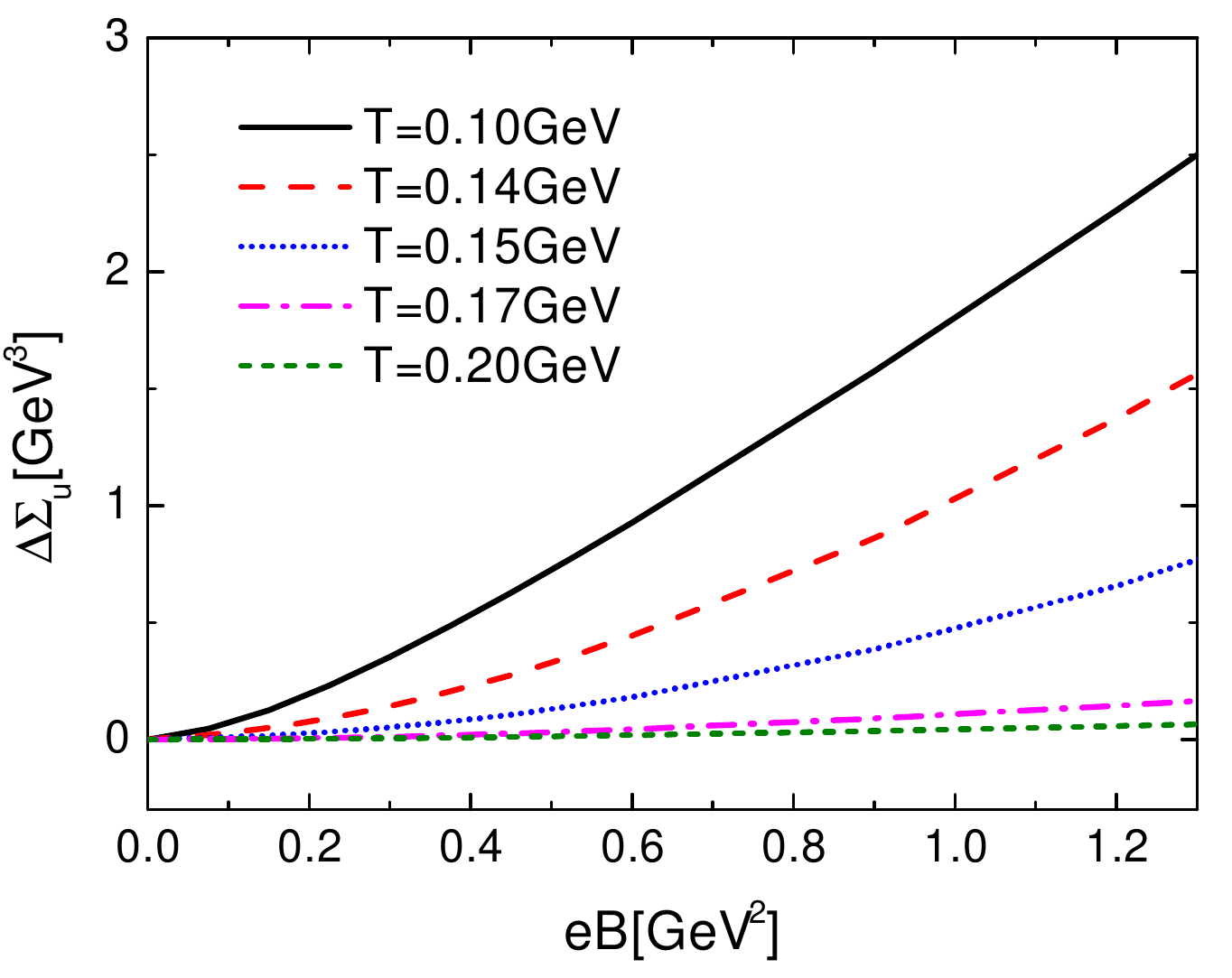}
    \includegraphics[width=0.9\columnwidth]{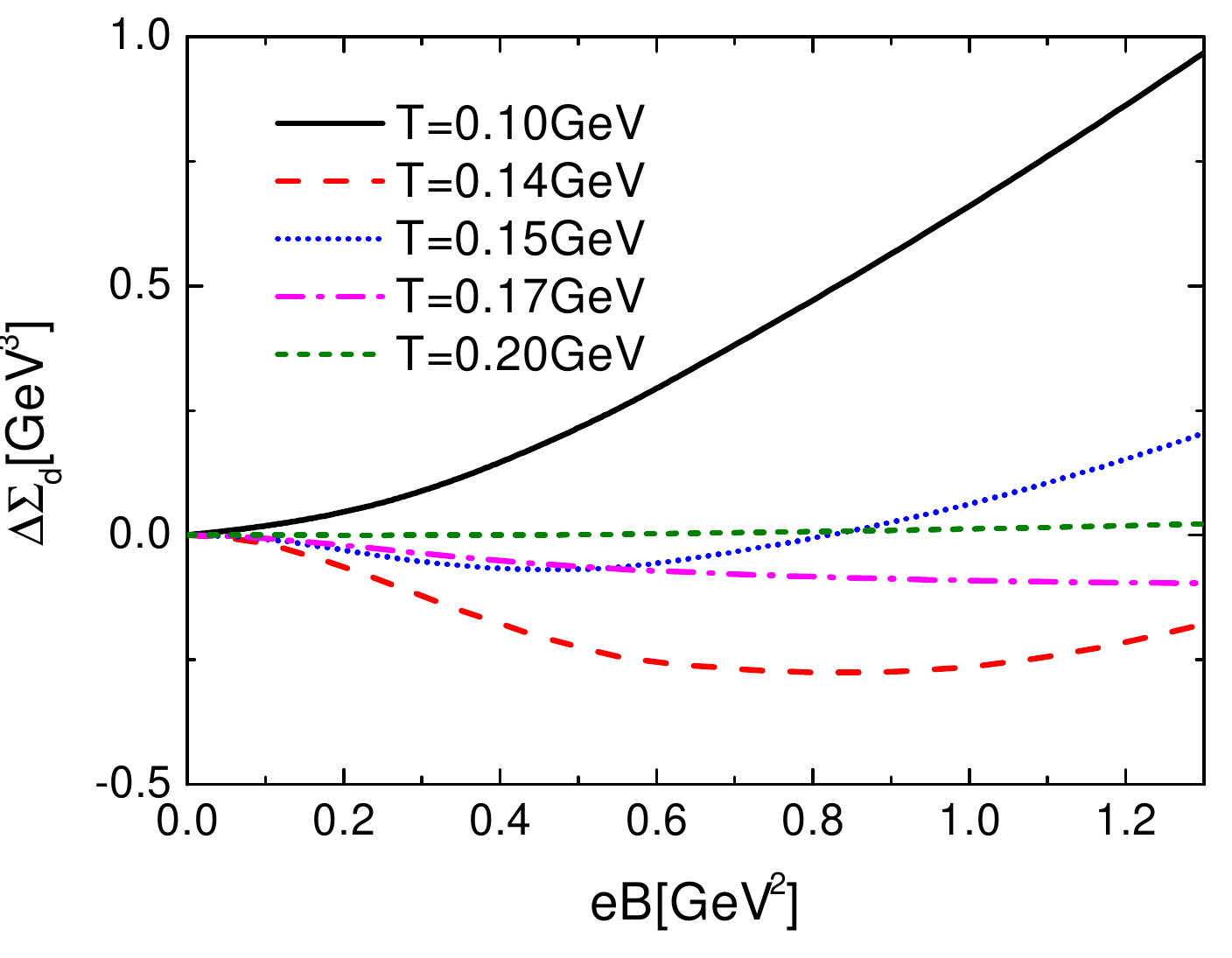}
  \caption{The magnetic field dependence of the chiral condensate at some temperatures,  for the averaged chiral condensate $\Delta{\Sigma}_{u/d}$ (\emph{left upper panel}) and the difference of up and down quark condensate $\Delta{\Sigma}_{u-d}$(\emph{right upper panel}). The  chiral condensate for up and down quark separately are  depicted  in the lower panel. }\label{fig:condTmu}
\end{figure*}

  Here by solving the coupled Dyson-Schwinger equations for quark and gluon propagator, we calculate the quark loop correction in the gluon propagator  self consistently.  We first depict the results for the gluon mass scale and the effective coupling in Fig.~\ref{fig:quarkloop}.   As the temperature  and the magnetic field strength increases,  the effective mass scale defined with  Eq.~\ref{eq:mg}  becomes larger.  Such an enhancement of mass scale translates into the suppression of quark gluon coupling as in the right panel of Fig.~\ref{fig:quarkloop}.  For low temperature, the magnetic field dependence of the coupling is also weak, and thus, the suppression of the coupling cannot compete with the enhancement of the quark mass function, which is the magnetic catalysis  phenomenon in the vacuum. However, for larger temperature, in particular in the regime close to the chiral phase transition, the coupling strength drops drastically. Meanwhile, the  up and down quark mass function is  small near phase transition, and thus the direct enhancement effect on quark mass  is also small.  In combination,   the  up and down quark mass function  is suppressed that yields  a quicker chiral restoration, which is the inverse magnetic catalysis phenomenon. Note that this competing  mechanism may fail for heavy quarks, where the  quark mass is too large for the suppression of the coupling to compete. 
  
After a complete scan over a wide range of temperatures and magnetic field strengths, we find that the chiral phase transition temperature indeed decreases upon incorporating the quark loop into the gluon propagator.   As depicted in Fig.~\ref{fig:Tc},   the phase transition temperature drops mildly from $T_c=152$ MeV at $eB=0$  to $T_c=142$ MeV for $eB\approx 1$ GeV$^2$.  In contrast, if one neglects the quark loop diagram in the gluon propagator, the quark mass function simply increases with the magnetic field strength, and the chiral phase transition temperature rises accordingly. This clearly demonstrates that inverse magnetic catalysis is driven by the enhancement of the gluon mass scale in the magnetic field.

\subsection{The chiral condensate as the function of temperature and magnetic field}

In Fig.~\ref{fig:condTmu}, we show the detailed results about the chiral condensate of up and down quark at some temperatures. For low temperatures, the averaged chiral condensate for up and down quarks increases as the temperature increases, following similar behavior in vacuum. As the temperature increases close to the chiral phase transition, the averaged chiral condensate drops and becomes negative at relatively large magnetic field, and moreover, it increases again for larger magnetic field. Above the chiral phase transition, even though the increasing and decreasing behavior becomes mild, the general behavior is similar to the case close to the chiral phase transition. 
For  extremely large  magnetic field, the quark mass may increase  for all the temperature, however, it is subtle to simply conclude it is the magnetic catalysis phenomenon. The increase of the condensate along the magnetic field is more drastic at low temperature, and hence, for large temperature, even though the condensate itself does not drop along the magnetic field, at a certain magnetic field its temperature dependence changes more quickly which still leads to a smaller phase transition temperature.

It is worth mentioning that such non-monotonic behavior arises solely from the down quark condensate, while for the up quark, the chiral condensate always increases along with  the magnetic field.  This is due to the difference in the electric charges of the up and down quarks: the up quark carries twice the electric charge of the down quark, leading to a much stronger response of the quark mass to the magnetic field, while the gluon propagator remains unaffected. Consequently, the chiral phase transition temperature for the up quark is barely altered, since the up quark condensate depends monotonically on the magnetic field. The decreasing phase transition temperature depicted in Fig.~\ref{fig:Tc} originates entirely from the decreasing behavior of the down quark condensate. This splitting behavior, however, raises a potential concern for the chiral phase transition temperature in extremely large magnetic fields: namely, there may exist a splitting of the chiral phase transition temperature between the up and down quarks, which deserves further careful investigation.
 
Besides, since the absolute value of the electric charge of the up quark is larger, the enhancement of the up quark mass function is generally stronger, and the up quark condensate remains larger than the down quark condensate in a finite magnetic field. At small magnetic field strengths, the difference between the up and down quark condensates decreases monotonically with increasing temperature.  This difference condensate becomes  non monotonous along with the temperature for  the magnetic field above $eB\sim 0.2$ GeV$^2$. This non-monotonic behavior, however, should be taken with a grain of salt, as it arises within the truncation scheme of our approach, which neglects the effect of the Polyakov loop potential in the quark propagator. A further study incorporating the Polyakov loop potential into the minimal QCD scheme is required to clarify this issue.
\section{Summary}\label{sec:sum}
It has been widely accepted that the magnetic field enhances chiral symmetry breaking by increasing the quark mass. This is understood as the magnetic catalysis phenomenon, which serves as the   baseline for the magnetic field effect on QCD matter. However, the quark loop contribution to the gluon propagator's self-energy allows the magnetic field effect to enter the gluon sector as well,   leading to an enhancement of the gluon screening mass. 

Using the Schwinger propagator formalism, one can readily generalize the Dyson–Schwinger equations for the quark and gluon propagators in the presence of a magnetic field. Upon solving these coupled equations, we find that at low temperatures with chiral symmetry breaking, the magnetic field effect on the quark mass function is dominant: as the magnetic field increases, the chiral condensate grows. Moreover, this enhancement is proportional to the quark mass itself. Therefore, at higher temperatures near the chiral phase transition, where chiral symmetry is close to being restored, the enhancement of the quark mass function becomes weak. Meanwhile, the enhancement of the gluon screening mass, induced by the magnetic field effect on the quark loop self-energy in the gluon propagator, is relatively insensitive to the quark mass and thus becomes dominant near the chiral phase transition. This enhancement of the gluon screening mass suppresses the QCD effective coupling and gives rise to inverse magnetic catalysis.

Furthermore, a detailed analysis reveals that the magnetic field effect on the up and down quarks is qualitatively different due to their different electric charges. The electric charge of the up quark is twice that of the down quark, leading to a much stronger response of the up quark mass function to the magnetic field: it always increases as the magnetic field strength grows. Consequently, inverse magnetic catalysis arises solely from the down quark condensate. This also suggests that for heavier quarks, inverse magnetic catalysis may disappear, as the quark mass enhancement in the magnetic field becomes too large to be overcome. In summary, we have provided a clear explanation of the magnetic field effects on the chiral phase transition. It is therefore of interest to pursue a further study incorporating the Polyakov loop potential into the current scheme, which would yield a more complete description of the QCD phase transitions in the presence of a magnetic field.

\begin{acknowledgments}
We thank Lang Yu, Shijun Mao, Defu Hou, Mei Huang and Jan M. Pawlowski for useful discussions.  We particularly thank Pengfei Zhuang, who has inspired and initiated this work after a discussion about the gluon mass and the effective coupling in the magnetic field. 
This work  is supported by the National Natural Science Foundation of China under Grants  No. 12305134, No. 12247107 and No. 12175007. 
\end{acknowledgments}

\bibliography{bibreferences}

\end{document}